# Dynamic symmetry breaking in chiral magnetic systems

J.A. Brock[1], M.D. Kitcher[2], P. Vallobra[1,3], R. Medapalli[1], M.P. Li[2], M. De Graef[2], S. Mangin[3], V. Sokalski[2], and E.E. Fullerton[1]

[1] Center for Memory and Recording Research, University of California, San Diego, La Jolla, CA USA

[2] Department of Materials Science and Engineering, Carnegie Mellon University, Pittsburgh, PA USA

[3] Université de Lorraine, CNRS, IJL, F-54000 Nancy, France

The Dzyaloshinskii-Moriya interaction (DMI) in magnetic systems stabilizes spin textures with preferred chirality, applicable to next-generation memory and computing architectures. In perpendicularly magnetized heavy-metal/ferromagnet films, the interfacial DMI originating from structural inversion asymmetry and strong spin-orbit coupling favors chiral Néel-type domain walls (DWs) whose energetics and mobility remain at issue. Here, we characterize a new effect in which domains expand unidirectionally from a combination of out-of-plane and in-plane magnetic fields, with the growth direction controlled by the in-plane field strength. These growth directionalities and symmetries with applied fields cannot be understood from static treatments alone. We theoretically demonstrate that perpendicular field torques stabilize steady-state magnetization profiles highly asymmetric in elastic energy, resulting in a dynamic symmetry breaking consistent with the experimental findings. This phenomenon sheds light on the mechanisms governing the dynamics of Néel-type DWs and expands the utility of field-driven DW motion to probe and control chiral DWs.

**Introduction**

In recent years, the impacts of the Dzyaloshinskii-Moriya interaction (DMI) in magnetic systems have attracted significant attention as the DMI can give rise to a spectrum of non-collinear spin textures, such as spin spirals, chiral domain walls (DWs), and magnetic skyrmions.[1] Dzyaloshinskii and Moriya first postulated the existence of this anti-symmetric exchange interaction in bulk non-centrosymmetric crystals.[2,3] Later, Fert predicted that an inherently tunable interfacial DMI (iDMI) exists in thin-film systems that feature interfaces between ferromagnets and heavy metals with large spin-orbit coupling.[4] Since then, there have been many reports detailing the engineering of DMI in materials systems.[5-8] Given the application potential for iDMI-stabilized skyrmions and chiral domain walls (DWs) in proposed racetrack-style memory architectures[9,10] and emerging neuromorphic computing approaches[11,12], the impact of the iDMI on the mobility and dynamics of spin textures is an active area of research. For sufficiently strong iDMI, magnetic films favor the formation of chiral Néel DWs over the achiral Bloch DWs expected in thin films with perpendicular anisotropy.[13,14] As Néel-type DWs are characterized by a magnetization rotation along the DW normal, an in-plane magnetic field breaks the azimuthal energy symmetry of the domain – leading to a difference in DW energies between the left and right side of a domain when an in-plane magnetic field is applied along a left-right axis.[15] Superficially, this energy asymmetry makes DW mobility on one side of a domain more favorable to reversal (in a direction collinear to the in-plane field), leading to an asymmetry in the field-induced domain expansion that can be used to identify the presence and magnitude of iDMI.[13,15-20]

Here, we report on a combined experimental and theoretical study of the magnetic reversal properties of Co/Ni/Pt-based thin-film multilayers with iDMI in response to both in- and out-of-plane magnetic fields. When imaging the reversal behavior, we observe domain growth directionalities that have not, as of yet, been reported. Specifically, we find that when the multilayers reverse via dendritic stripe domains, the domain growth is highly anisotropic, and the growth direction changes dramatically depending on the strength of the in-plane field. Furthermore, the growth direction symmetries observed when changing the sign of the in- and out-of-plane fields requires a breaking of time-reversal mirror symmetry along both the horizontal and vertical planes that cannot be explained from static energies alone. We posit that the observed domain growth behaviors arise from a dynamic symmetry-breaking during domain expansion. Our calculations of the steady-state dynamical reorientation of the DW magnetization profile and the associated changes in the dispersive stiffness permit a quantitative understanding the experimentally observed behaviors.

**Experimental Results**

Hysteresis loops for samples with a varying number of [Co(0.7 nm)/Ni (0.5 nm)/Pt (0.7 nm)] repetitions $N$ are shown in Figure S1 in the Supporting Information. All samples exhibit strong perpendicular magnetic anisotropy, characterized by a full remanence at zero field in the perpendicular measurement geometry. Loops collected in the in-plane geometry demonstrate that the in-plane saturation field $\mu_0 H_S$ does not change significantly with $N$. Given that all samples exhibit a similar $\mu_0 H_S$ (~ 1 T) and saturation magnetization $M_S$ (~ 1000 kA/m), the effective perpendicular magnetic anisotropy energy density $K_{eff}$ (5 x $10^5$ J/m$^3$) is relatively invariant within $1 \leq N \leq 5$.

Polar Magneto-optic Kerr effect (MOKE) images of the $N = 2$ sample shown in Figure 1a-d illustrate that this sample reverses via large, nearly circular domains. Figure 1a shows that the domains expand isotropically absent an energy symmetry-breaking in-plane magnetic field. In agreement with previous reports on systems with iDMI[15-22], a mild growth asymmetry is present when an in-plane magnetic field is applied, and the favored growth direction is antiparallel to $\mu_0 H_x$ (Figure 1b-c). Knowing the perpendicular orientation of the domain, this is consistent with a left-handed Néel chirality, as has been observed in other Pt/Co/Ni-type systems.[17,18] By measuring the growth velocity as a function of $\mu_0 H_x$ for both the left and right sides of the domain, we estimate an iDMI energy density $D$ of -0.63 mJ/m$^2$ for the [Co/Ni/Pt]$_2$ sample (see Section S2 of the Supporting Information). Throughout this work, a negative value of $D$ denotes a left-handed chirality.

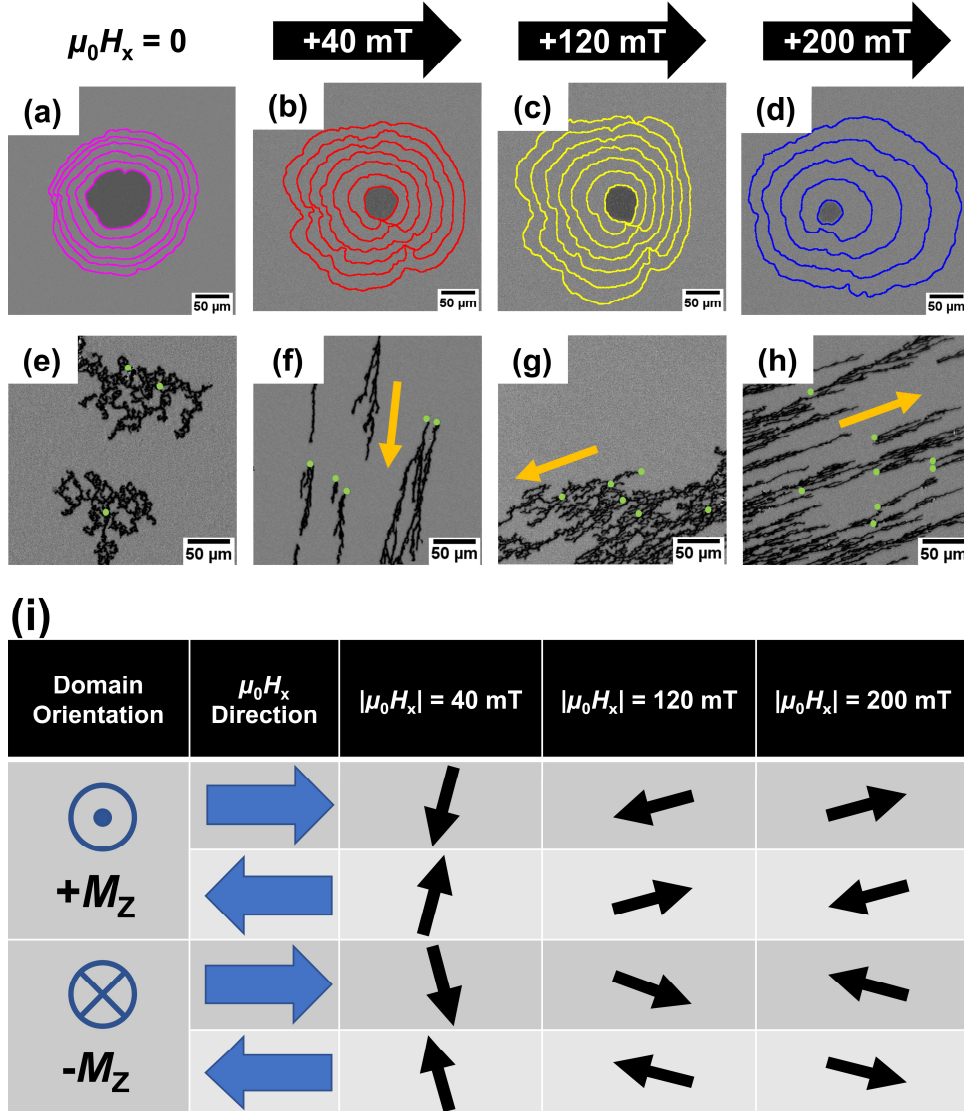

**Figure 1.** Polar MOKE images of magnetic domain growth in the [Co/Ni/Pt]$_2$ (a-d) and [Co/Ni/Pt]$_3$ (e-h) samples in response to 5 ms-long 15 mT out-of-plane magnetic field ($\mu_0 H_z$) pulses collected in static in-plane magnetic fields $\mu_0 H_x$ of 0 (a, e), +40 mT (b, f), +120 mT (c, g), and +200 mT (d,h). (i) Favored growth direction of dendritic stripe domains in the [Co/Ni/Pt]$_3$ sample for different permutations in domain orientation and in-plane field $\mu_0 H_x$ magnitudes and directions. Each colored ring in (a-d) represents the extent of the domain after three $\mu_0 H_z$ pulses

were applied. The images in (e-h) were collected after three magnetic field pulses were applied, while the green markers represent the initial nucleation sites of reverse domains.

For higher-magnitude in-plane fields, the favored growth direction reverses, as shown in Figure 1d. There have been similar reports of the high-energy side of a bubble domain growing faster than its low-energy counterpart above a threshold in-plane field strength.[18,21,22] Subsequently, it was analytically and micromagnetically demonstrated that in systems with iDMI, the elastic energy of the DW (which governs motion in the creep regime) differs significantly from the linear wall energy generally employed in DW creep models.[18] As such, this behavior in the [Co/Ni/Pt]$_2$ sample can be explained within the context of surface stiffness in the long-wavelength regime, as shown in Section S3 of the Supporting Information.

In agreement with the quasi-static magnetometry data, the samples with $N \geq 3$ reverse via dendritic stripe domains, as shown for the [Co/Ni/Pt]$_3$ sample in Figure 1e-h. Much like the circular domains in the samples with $N = 1$ or 2, the growth of these ~3 µm-wide stripe domains is nearly isotropic when $\mu_0 H_x = 0$, as shown in Figure 1e. The green dots show the initial nucleation sites of the dendritic domains in Figure 1e-h. This lack of a favored growth direction persists until $\mu_0 H_x = + 40$ mT, at which point the domains exhibit a strong directionality to their growth, as opposed to the mild asymmetry seen for the same field in the circular domains; that is, once nucleated, the domains prefer to grow in one specific direction, as indicated by the arrows in Figure 1f-h. Surprisingly, the favored growth direction is nearly perpendicular to the $\mu_0 H_x$ axis for modest in-plane fields – growth behavior that would not be expected from the chiral Néel-type DWs favored by the iDMI. From Lorentz TEM images of [Co/Ni/Pt]$_5$ multilayers collected in the absence of applied fields (Figure S2 in the Supporting Information), a preference for Néel-type DWs exists for these types of samples. As shown in Figure 1g, when $\mu_0 H_x$ is increased to +120 mT, the growth direction becomes more collinear to the $\mu_0 H_x$ axis in a manner more consistent with the left-handed chirality inferred in the [Co/Ni/Pt]$_2$ sample. However, when $\mu_0 H_x$ is increased above +140 mT, a rapid ~180° change in the growth direction occurs, as shown in Figure 1h for $\mu_0 H_x = +200$ mT. The favored growth direction of dendritic stripe domains in the [Co/Ni/Pt]$_3$ sample for various permutations in domain orientation and $\mu_0 H_x$ magnitudes/directions are summarized in Figure 1i; a version of this figure where MOKE images have replaced the schematic arrows is provided in Figure S3 in the Supporting Information.

In Figure 2a, we plot the growth direction $\theta_{\text{growth}}$ of "up" ($+M_z$) domains as a function of in-plane field strength $\mu_0 H_x$ using the $\theta_{\text{growth}}$ and $\mu_0 H_x$ conventions defined in Figure 2b. A plot of $\theta_{\text{growth}}(\mu_0 H_x)$ for both perpendicular domain orientations is provided in Figure S4 of the Supporting Information. While an analytical model elucidating these trends is presented later in this manuscript, we first discuss some of the noteworthy features of domain growth demonstrated in Figure 1 and Figure 2. First, we note that for a fixed perpendicular orientation of reverse domains and $\mu_0 H_x$ magnitude, changing the $\mu_0 H_x$ direction by 180° also changes $\theta_{\text{growth}}$ by 180°, as would be expected in a chiral magnetic system.[15]

However, when considering the expansion of domains with opposite perpendicular orientations, we find that the $\hat{x}$-component of the growth direction changes while the $\hat{y}$-component (vertical direction in the images) is invariant. This peculiar behavior is particularly apparent in the limit of $|\mu_0 H_x| = 40$ mT, where $+M_z$ and $-M_z$ domains grow in the same direction, nearly vertical to the $\mu_0 H_x$ axis. Moreover, this vertical asymmetry changes orientation upon inversion of the $\mu_0 H_x$ direction. These surprising observations are not commensurate with Bloch- or Néel-type chiral DWs in which the chirality is set by intrinsic energetic factors. Besides these trends to the $\theta_{growth}(\mu_0 H_x)$ profiles, the directional growth characteristics of the dendritic stripe domains are independent of the in-plane field history, as the in-plane field strength and direction can be changed during the reversal process, and the growth direction varies "on-demand" (Figure S5 of the Supporting Information). This history-independence indicates that the energy landscape governing the observed growth asymmetries is not set during the initial nucleation of a domain. Additionally, we note that similar growth behaviors are seen in samples with $N = 4$ or 5, and the results are insensitive to modest changes in the Co, Ni, or Pt thicknesses.

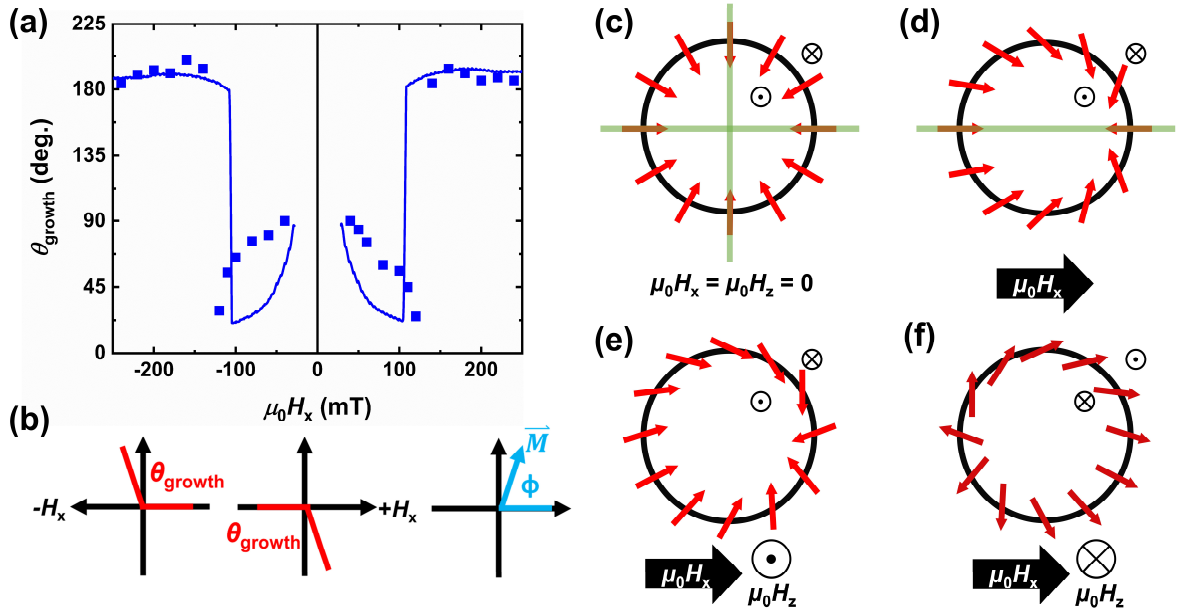

**Figure 2: (a)** The growth direction $\theta_{growth}$ of "up" ($+M_z$) dendritic domains as a function of $\mu_0 H_x$, determined both experimentally (symbols) and analytically with steady-state dynamics and dispersive stiffness (lines). When plotting the analytically predicted $\theta_{growth}$ profiles, we only display data for in-plane fields where a > 60 % growth asymmetry between the fastest and slowest moving azimuthal positions is predicted (*i.e.*, $|\mu_0 H_x| \geq 40$ mT). **(b)** The conventions used throughout this work to define the domain growth direction $\theta_{growth}$ relative to the in-plane field $\mu_0 H_x$ direction and the DW core magnetization angle $\varphi$. **(c-f)** Schematic depictions of DW magnetization profiles in **(c)** zero applied magnetic field, **(d)** an in-plane magnetic field, **(e)** an in-plane and out-of-plane magnetic field, and **(f)** an in-plane and in-to-plane magnetic field. Green lines in **(c-d)** indicate planes of time-reversal mirror symmetry.

Secondly, there is a strong selectivity to the growth direction of dendritic stripe domains (Figure 1f-h), compared to the mild growth asymmetries seen in the samples that reverse via large circular domains (Figure 1b-d). In films with perpendicular anisotropy, dendritic stripe domains form out of a balance between ferromagnetic exchange and long-range dipolar fields. The energetic impetus for the formation of different types of domain morphology has been extensively studied in films without iDMI, starting with the work of Kittel in the 1940s.[23-25] In general, the equilibrium domain size decreases exponentially with increasing film thickness in the thin-film limit before reaching a minimum at intermediate thickness, followed by an increase in size proportional to the square root of the film thickness in thicker films. While the iDMI is known to lead to smaller equilibrium domain sizes, a similar relationship between film thickness and equilibrium domain size holds in films with this additional energy term.[26] For the [Co/Ni/Pt] films discussed here, we observe a transition from relatively large domains to stripe domains when increasing the number of repetitions $N$ from 2 to 3 (*i.e.*, within the thin film limit, where the equilibrium domain size decreases exponentially with increasing film thickness). Magnetic reversal via dendritic stripe domains occurs by means of growth from a nucleation site, as the stripe tip curvature (and thus, the stripe width) stays more or less constant while the length of the stripe domains increases to fill the area of the film.[27] This reversal phenomenon is markedly different from thinner films with larger domains, where domains tend to uniformly expand in all directions, filling the area of the film to obtain a single reverse domain state. In concert with the localized chirality of the DW magnetization profile (and, hence, the localized energy profile) at the tips of the domains[28], the dendritic domain morphology allows for a stronger link between the azimuthal energy asymmetry and the expansion asymmetry. As we discuss in the subsequent section, this leads to the larger-aspect ratio domains and stronger directional growth observed in the [Co/Ni/Pt]$_3$ sample reversing via stripe domains compared to the thinner samples.

**Analytical Modeling**

As shown in Figure S6 of the Supporting Information, domain growth in the [Co/Ni/Pt]$_3$ sample falls within the thermally activated creep regime for the $\mu_0 H_z$ and $\mu_0 H_x$ fields employed experimentally throughout this work.[15,29-31] Within the creep regime, the velocity of domain expansion $v$ follows an Arrhenius scaling law, given by:[29,32,33]

$$v = v_0 \cdot e^{-\chi \left(\frac{\varepsilon}{H_z}\right)^{1/4}} \tag{1}$$

where $v_0$ is the velocity scaling factor, $\chi$ is an energy barrier scaling constant, $H_z$ is the perpendicular field strength, and $\varepsilon$ is the elastic energy of the DW. A defining hallmark of Equation 1 is that $v$ scales inversely with $\varepsilon$. In the case of magnetic domains, $\varepsilon$ is generally taken as the 1D linear DW energy density $\sigma$, expressed as:[18]

$$\sigma[\theta, \varphi] = \sigma_0 + (\ln 2) t_f \pi^{-1} \mu_0 M_s^2 \cos^2[\varphi - \theta] - \pi D \cos[\varphi - \theta] - \pi \lambda_0 \mu_0 M_s H_x \cos \varphi \tag{2}$$

In Equation 2, $\theta$ is the azimuthal orientation of a given DW section and $\varphi$ is the DW core magnetization angle, defined using the conventions shown in Figure 2b. The first term of Equation 2 is the resting energy of a Bloch-type DW ($\sigma_0 = 4\sqrt{AK_{eff}}$). The second term represents the DW anisotropy energy, which increases with film thickness ($t_f$) and degenerately favors Bloch DWs of either chirality. The third term gives the contribution of the effective field due to the iDMI, which stabilizes Néel DWs of the handedness determined by the sign of $D$. The final term is the Zeeman energy emerging from the application of an in-plane magnetic field along the *x*-axis. In Equation 2, $\mu_0$ is the vacuum permeability and $\lambda_0 = \sqrt{A/K_{eff}}$ (where $A$ is the exchange stiffness) is the DW width. To obtain the resting equilibrium energy and in-plane magnetization of a DW segment with azimuthal orientation $\theta$, the energy functional is minimized with respect to $\varphi$. As shown in Figure 2c, when $\mu_0 H_x$ is zero, Equation 1 predicts a DW magnetization profile — and hence, an energetic profile — that has time-reversal mirror symmetry about the horizontal and vertical axes of the domain.

In Figure 3a-d, we show the static equilibrium DW magnetization profiles $\varphi(\theta)$ predicted by Equation 2 for the [Co/Ni/Pt]$_3$ sample for various $\mu_0 H_x$ values, using the sample properties listed in the caption of Figure 3. When $\mu_0 H_x$ = 0, the DWs are predicted to exhibit a significant Néel character despite the higher DW anisotropy expected in samples with a higher number of [Co/Ni/Pt] repetitions — a prediction confirmed from the Lorentz TEM images shown in Figure S2 of the Supporting Information. As the magnitude of $\mu_0 H_x$ is increased, the DW magnetization profile is progressively pulled in the direction of $\mu_0 H_x$. From the static equilibrium treatment of the DW magnetization profile, growth asymmetries collinear to the in-plane field axis are the only expected outcomes, as the time-reversal mirror symmetry of the DW profile is only broken about the vertical axis (Figure 2d).

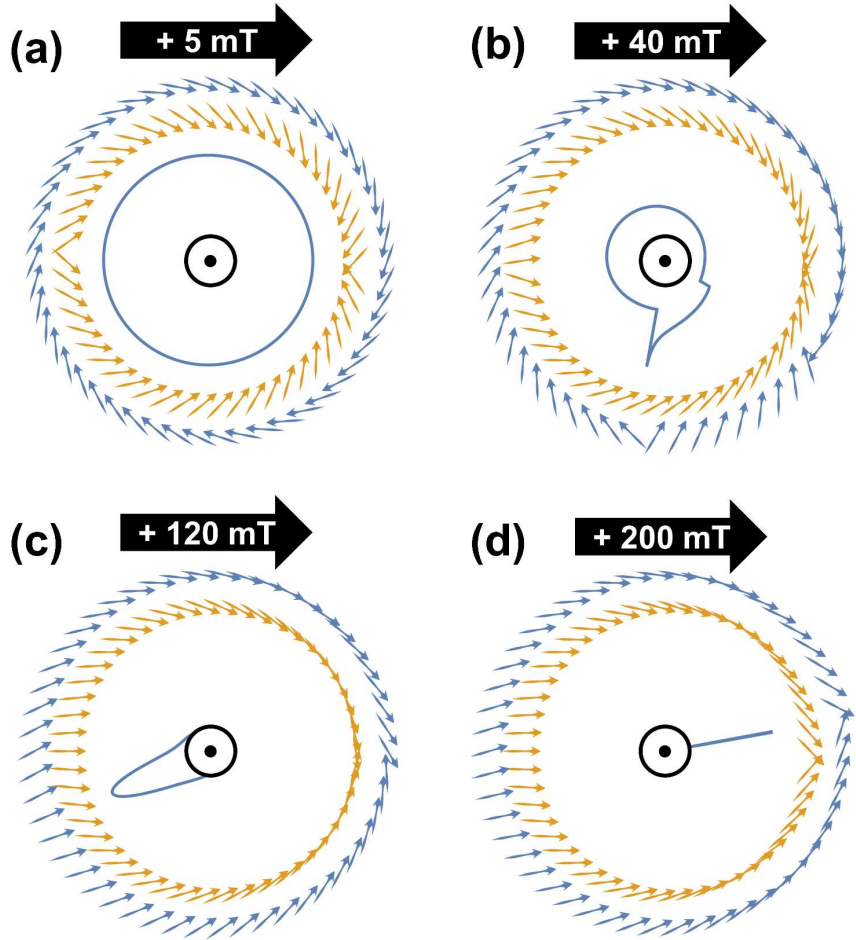

**Figure 3.** Predicted DW magnetization profiles in the [Co (0.7 nm)/ Ni (0.5 nm)/ Pt (0.7 nm)]$_3$ sample under the assumption of static equilibrium (gold arrows) and steady-state dynamic reorientation (blue arrows) under applied in-plane fields of (a) $\mu_0 H_x$ = +5, (b) +40 mT, (c) +120 mT, and (d) +200 mT. The polar profile shown in the center of each diagram represents the relative velocity of expansion ($v/v_{max}$) as a function of azimuthal position predicted from Equation 4. When using Equation 2 and Equation 3 to calculate the static equilibrium and steady-state dynamic profiles (respectively), the sample properties were as follows: $M_S$ = 1000 kA/m, $K_{eff}$ = 5 x 10$^5$ J/m$^3$, $t_f$ = 3.6 nm, $D_{DMI}$ = -0.45 mJ/m$^2$, $\alpha$ = 0.208, and $\mu_0 H_z$ = +15 mT.

In addition to the static energies reflected in Equation 2, when a perpendicular field $\mu_0 H_z$ is used to propagate a reversed magnetic domain, a torque will act on the internal magnetization of the DW, rotating $\varphi(\theta)$ away from the static equilibrium profile predicted by Equation 2. This field torque will be counteracted by the balance of the effective fields, also captured by Equation 2. While large perpendicular fields will cause a DW's core magnetization to precess[34], a DW within the creep regime is expected to experience a steady-state dynamical reorientation of its internal magnetization, described within the Landau-Lifshitz-Gilbert framework by:[35]

$$\dot{\varphi} = \frac{\gamma\mu_0}{1+\alpha^2}[-\alpha\Omega_A + H_z] = \frac{\gamma\mu_0}{1+\alpha^2}\left[-\frac{\alpha\sigma_\varphi}{2\lambda_0\mu_0 M_S} + H_z\right] \quad (3)$$

where $\gamma$ is the gyromagnetic ratio, and $\alpha$ is the Gilbert damping parameter. In Equation 3 and henceforth, subscripts to the DW energy density $\sigma$ denote a partial derivative. The balance of these competing factors can result in a steady-state $\varphi(\theta)$ profile that differs from the static equilibrium case, which in turn modifies the relevant energy-related terms. From the schematic depictions of the effect of this torque on chiral Néel-type walls (Figure 2e-f), it can be surmised that it acts in the same rotational sense for both $+M_z$ and $-M_z$ domains – akin to the promotion of Bloch-type walls of opposite chirality depending on the orientation of the reversal domain/sign of $\mu_0 H_z$, thus breaking the mirror symmetry of the DW magnetization profile along both the horizontal and vertical axes.

Recently, Pellegren *et al.* showed that when DW energies are anisotropic in $\theta$, the elastic energy scale $\varepsilon$ is no longer given simply by the 1D static DW energy $\sigma$, but by the dispersive stiffness.[18] As the dispersive stiffness accounts for the pinning length scale and the local curvature of the energy landscape through additional energy terms, a more accurate picture of the elastic behavior of DWs can be obtained. For a DW of arbitrary length $L$, we employ the generalized dispersive stiffness $\tilde{\sigma}$ to fully capture the steady-state reorientation using the expression:

$$\tilde{\sigma}[\theta, \varphi, L] = \sigma + \sigma_{\theta\theta} - \frac{\sigma_{\varphi\theta}^2}{\sigma_{\varphi\varphi}}\zeta\left[\frac{L}{2\Lambda}\right] \quad (4)$$

where $\zeta(l) = 1 - \frac{3}{l^2}(l - \tanh(l))$ and $\Lambda = \lambda_0\sqrt{\sigma_0/\sigma_{\varphi\varphi}}$. In Equation 4, all derivatives are evaluated at the steady-state magnetic orientation $\varphi$ corresponding to each $\theta$. The second and third terms of Equation 4 represent the curvature contributions that emerge due to anisotropy in the DW energy with respect to both $\theta$ and $\varphi$. The third term expresses the dependence of the stiffness on $L$ and the exchange length scale $\Lambda$; for finite values of $L$, we recognize that for vanishing values of $\sigma_{\varphi\varphi}$, $\lim_{\sigma_{\varphi\varphi}\to 0}\frac{\sigma_{\varphi\theta}^2}{\sigma_{\varphi\varphi}}\zeta\left[\frac{L}{2\Lambda}\right] = -\frac{\sigma_{\varphi\theta}^2 L^2}{10\sigma_0\lambda^2}$. Consequently, by calculating the dispersive stiffness in the steady-state configuration as a function of $\theta$, the relative velocity of a wall segment at all azimuthal orientations can be computed, and the most favorable growth directions of a domain can be determined. For $\theta$ positions where $\tilde{\sigma} < 0$, the DW magnetization profile is expected to facet, adopting the orientation and magnetization of the neighboring segments (as predicted by the Wulff construction), driving the $\tilde{\sigma}$ values to zero.[18]

Defining $\theta_{growth}$ as the azimuthal position having the lowest $\tilde{\sigma}$ value, we obtain the most suitable match between the experimental and analytically predicted $\theta_{growth}(\mu_0 H_x)$ profiles when the parameters $D = -0.45$ mJ/m$^2$, $\alpha = 0.208$, and $L = 70$ nm are used in the modeling, along with $A_{ex} = 10$ pJ/m and the experimentally determined $M_S$ and $K_{eff}$ values, as shown in Figure 2a. While $L = 70$ nm was used in the present calculations, a similar degree of agreement between the experimental data and the modeling was found over a wide range of $L$ values. Also, when the experimentally-determined $D$ value of $-0.63$ mJ/m$^2$ is used, qualitatively-similar results are obtained, with a modest offset to the characteristic $\mu_0 H_x$ at which the ~180° change in growth direction occurs. In Figure 4a-d, we show the dispersive stiffness $\tilde{\sigma}$ and predicted relative velocity of expansion $v/v_{\max}$ as a function of azimuthal angle $\theta$ for the [Co/Ni/Pt]$_3$ sample in several in-plane fields $\mu_0 H_x$; an alternative depiction of the azimuthal dependence of the velocity is provided in Figure 3. $\tilde{\sigma}$ is determined using the approach discussed above, whereas $v/v_{\max}$ was determined using a modified expression for the creep velocity that takes the dispersive stiffness into account, given as:[18]

$$v = v_0 \cdot \exp\left[-\frac{\kappa \cdot (\tilde{\sigma}[H_x])^{1/4}}{H_z \cdot (\tilde{\sigma}[H_x=0])^{1/4}}\right] \tag{5}$$

where the velocity scaling ($v_0$) and energy barrier scaling ($\kappa$) parameters are determined from fitting to the experimental $v(\mu_0 H_x, \mu_0 H_z)$ data, as demonstrated in Figure S6 of the Supporting Information.

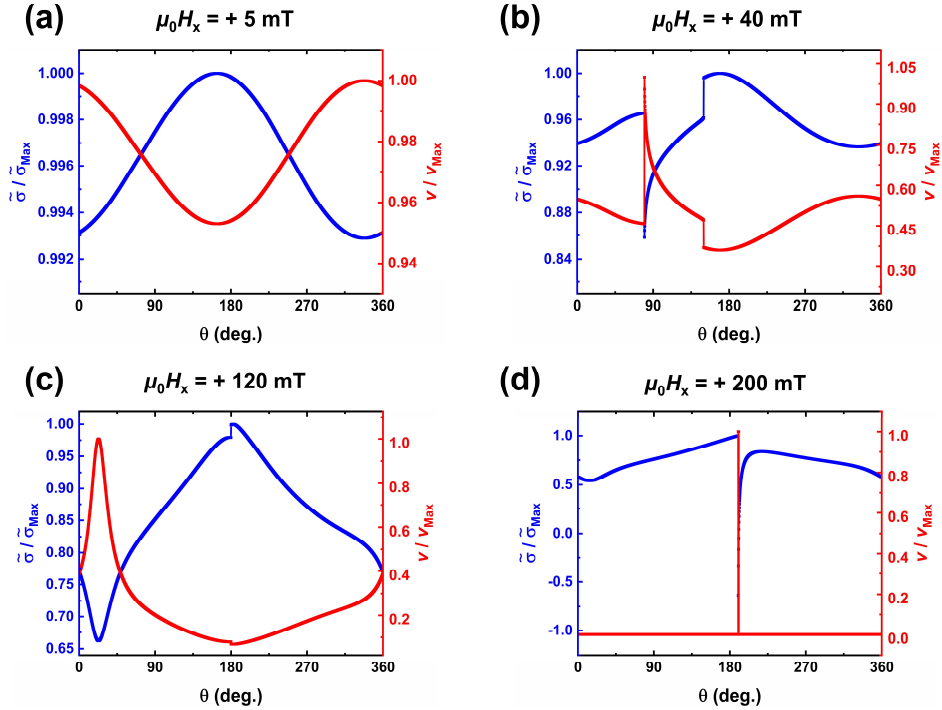

**Figure 4.** Dispersive stiffness $\tilde{\sigma}/\tilde{\sigma}_{\max}$ (blue lines) and predicted relative expansion velocity $v/v_{\max}$ (red lines) as a function of azimuthal position $\theta$ for $+M_z$ dendritic stripe domains in the [Co/Ni/Pt]$_3$ sample in applied in-plane fields $\mu_0 H_x$ of (a) +5 mT, (b) +40 mT, (c) +120 mT, and (d) +200 mT. Throughout Figure 4, $\theta$ is defined using the angular convention for a positive $\mu_0 H_x$ shown in Figure 2b, and $L$ is 70 nm.

For low magnitude $\mu_0 H_x$ values (Figure 4a), $\tilde{\sigma}$ is rather invariant with $\theta$; consequentially, the velocity asymmetry with respect to $\theta$ is muted. However, as $\mu_0 H_x$ is increased to + 40 mT, the $\tilde{\sigma}(\theta)$ and $v(\theta)$ profiles become substantially more anisotropic (Figure 4b), favoring domain growth in a direction similar to that observed experimentally. While the degree of growth asymmetry seen experimentally when $\mu_0 H_x$ = +40 mT is more heightened than what is predicted in Figure 4b, we believe differences in the DW width, $L$, and/or field-induced pinning not accounted for in the modeling strengthen the preference for growth in the experimentally observed direction. We also note that under the long-wavelength limit of the stiffness model (*i.e.*, $L \to \infty$) for 35 mT < $|\mu_0 H_x|$ <

70 mT, the calculated stiffness profiles strongly favor growth in both the experimentally observed vertical direction, as well as in a direction collinear to the applied $\mu_0 H_x$ (Figure S7 in the Supporting Information); growth along the latter direction could be suppressed for the same reasons mentioned above. Nevertheless, our analytical modeling explains how both horizontal and, more crucially, vertical asymmetries can develop in the magnetization profile, giving rise to the observed growth behaviors. Additionally, the model correctly predicts the symmetries reflected in Figure 1i and Figure S3 in the Supporting Information – namely, that the vertical growth direction under a given in-plane field does not change with the domain polarity. Coupled together, these two traits of our modeling punctuate the critical role that dynamic symmetry breaking plays in describing our experimental results.

Further increasing $\mu_0 H_x$ to +120 mT (Figure 4c) and +200 mT (Figure 4d), a strong correspondence is maintained between the experimental and predicted $\theta_{growth}(\mu_0 H_x)$ and $v(\theta)$ behavior. As shown in Figure 3, the $\theta$ positions of minimal $\tilde{\sigma}$ correspond to regions where the $\varphi(\theta)$ profile exhibits discontinuities (short of the full $2\pi$-rotations seen at Bloch points in DWs), where the local energy and net restoring torque approach an inflection point and a local maximum, respectively (Figure S8 in the Supporting Information). As previously reported by Sanchez-Tejerina *et al.*, this abrupt reorientation of the magnetization profile stems from variations in the driving and restoring torques acting on the DW, which arise from the $\theta$-dependent competition between the effects of $\mu_0 H_z$, $\mu_0 H_x$, iDMI, and DW anisotropy.[33] The highly localized nature of these minima in $\tilde{\sigma}$ - combined with the heightened configurational sensitivity of dendritic stripe domains - enables the strong directionality to the domain growth observed.

**Conclusion**

We report the in-plane field-induced magnetic domain growth asymmetries present in several Co/Ni/Pt–based thin film heterostructures engineered to possess an interfacial Dzyaloshinskii-Moriya interaction that favors chiral Néel-type DWs. For thinner samples that reverse their magnetization via large circular domains, growth asymmetries collinear to the symmetry-breaking in-plane magnetic field are observed, as predicted by the current understanding of the interfacial Dzyaloshinskii-Moriya interaction within the typical long-wavelength stiffness model of the elastic energy of a DW. For thicker samples, where reversal occurs via dendritic stripe domains, anomalous growth directionalities with respect to the in-plane field strength and direction emerge – most notably, growth perpendicular to low-magnitude in-plane fields. Through an analytic treatment of steady-state dynamical symmetry breaking and the dispersive stiffness of DWs, it is possible to model the magnetization configurations that would give rise to this unexpected growth behavior and its symmetries with respect to the out-of-plane field direction. These results shed light on the mechanisms governing the dynamics of Néel-type DWs and provide new opportunities for the control of chiral magnetic systems.

**Experimental Methods**

Multilayer stacks of the structure Ta(3)/Pt(3)/[Co(0.7)/Ni(0.5)/Pt(0.7)]$_N$/Ta(3) (thicknesses in nm) were grown on Si substrates with a 300 nm-thick thermal oxide (SiO$_x$) coating by dc magnetron sputtering, using a power of 50 W in a 3 mTorr Ar pressure. The number of [Co/Ni/Pt] repetitions $N$ was varied between 1 and 5. Our choice of materials was motivated by reports that Pt/Co and Pt/Ni interfaces are predicted to have different signs of iDMI[36], such that an additive iDMI may be achieved when they are incorporated in a structurally asymmetric Pt/Co/Ni/Pt heterostructure.[37]

The static magnetic properties were assessed using the vibrating sample magnetometry (VSM) technique at room temperature in both the in-plane and perpendicular geometries. To image the directionality of domain growth, we used a MOKE imaging platform manufactured by Evico Magnetics, configured for polar MOKE sensitivity. Perpendicular magnetic field pulses that were 5 ms-long and 15 mT-strong were used to nucleate and expand the magnetic domains. The static in-plane magnetic fields were varied in magnitude and direction between 0 and 320 mT. As is standard for MOKE imaging, the micrographs presented herein were collected using a background subtraction of the signal when the sample magnetization was fully saturated. We define a positive in-plane magnetic field as pointing along the positive $x$-axis, as indicated in Figure 2b. Unless otherwise stated, the sample magnetization was saturated opposite to the intended perpendicular field pulse direction between trials. Fresnel-mode Lorentz transmission electron microscopy (LTEM) was performed on an aberration-corrected FEI Titan G2 80-300 operated in Lorentz mode at room temperature. Samples for LTEM measurements were grown on 30 nm-thick SiN windows.

**Acknowledgements**

J.A.B., R.M., and E.E.F. acknowledge support through the QMEEN-C Energy Frontier Research Center, funded by the U.S. Department of Energy, Office of Science (Award No. DE-SC0019273) and from the University of California Office of the President through the Multicampus Research Initiatives Program on Electrical Control of Topological Magnetic Order. M.D.K., M.P.L., M.D.G., and V.S. acknowledge support from the Defense Advanced Research Agency (DARPA) program on Topological Excitations in Electronics (TEE) (Grant No. D18AP00011) and use of the Materials Characterization Facility at Carnegie Mellon University supported by grant MCF-677785. M.D.K. also acknowledges support from the National GEM Consortium, as well as the Neil and Jo Bushnell Fellowship in Engineering from the College of Engineering at Carnegie Mellon University. The authors are grateful to Mark van Ommeren for assistance with MOKE imaging in the initial stage of this project.

**Supporting Information:**

**Section S1. Supplemental Figures**

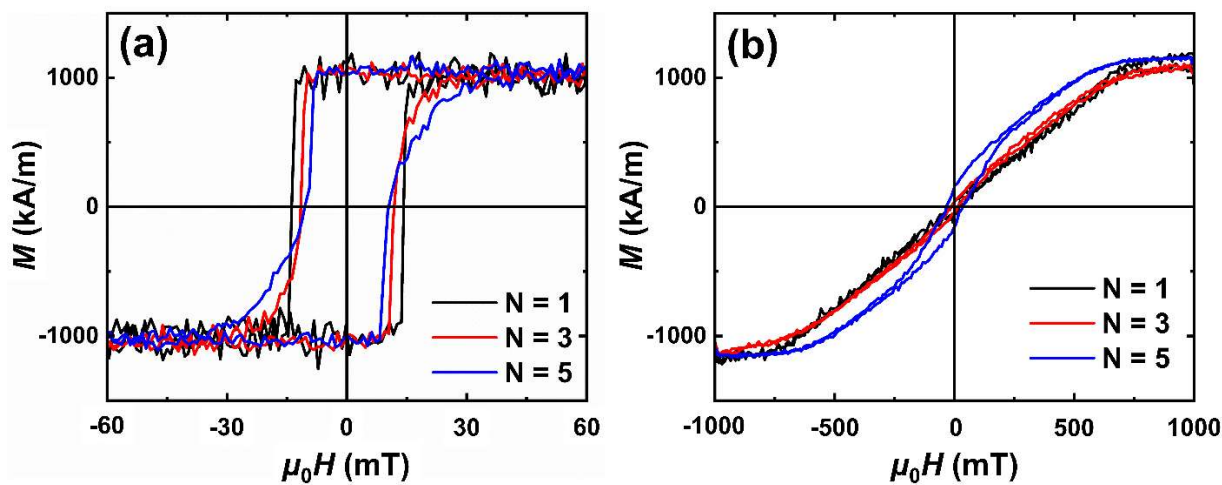

**Figure S1:** Room temperature magnetic hysteresis loops of the [Co(0.7 nm)/Ni (0.5 nm)/Pt(0.7 nm)]$_N$ samples collected in the (a) out-of-plane and (b) in-plane geometries using vibrating sample magnetometry.

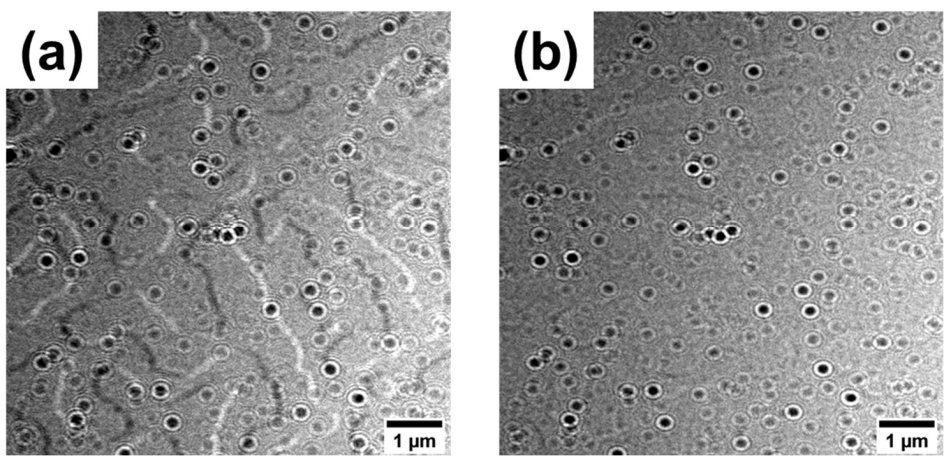

**Figure S2:** Lorentz TEM images of the [Co(0.7 nm)/Ni(0.5 nm)/Pt(0.7 nm)]$_5$ sample, collected in zero applied magnetic fields when the sample was tilted (a) 30° and (b) 0° relative to the imaging beam. A domain state was formed before imaging through the application of a perpendicular magnetic field. The circular black features correspond to voids in the sample and/or debris on the surface.

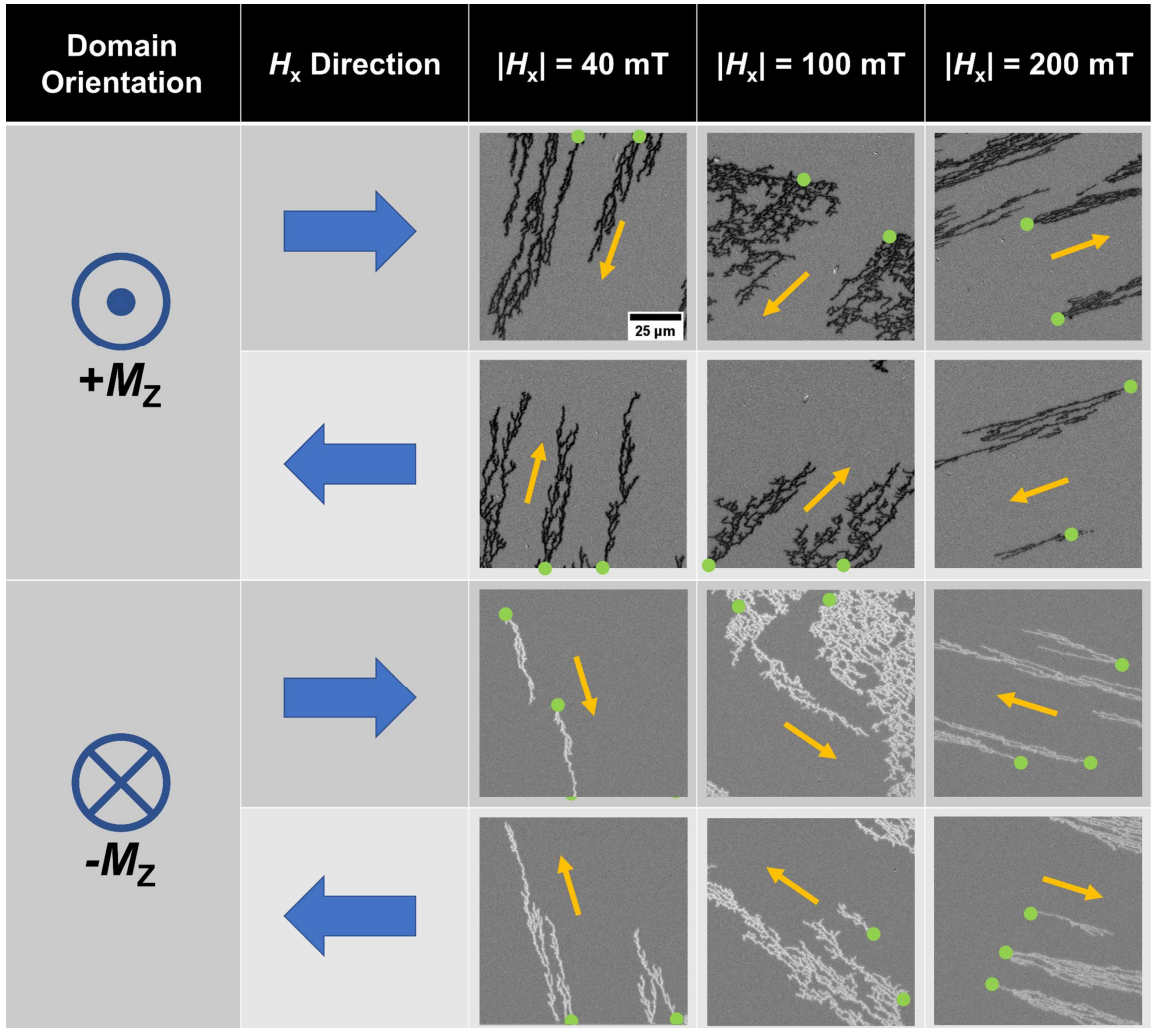

**Figure S3:** MOKE images depicting the growth of dendritic stripe domains in the [Co(0.7 nm)/Ni(0.5 nm)/Pt(0.7 nm)]$_3$ sample for different permutations of domain orientation and in-plane field $\mu_0 H_x$ magnitudes and directions. The images were collected as 5 ms-long, 15 mT-strong out-of-plane magnetic field pulses were applied. Orange arrows indicate the favored growth direction. Green dots in the images indicate the initial nucleation sites; dots at the edge of the images indicate a nucleation site out of the field of view.

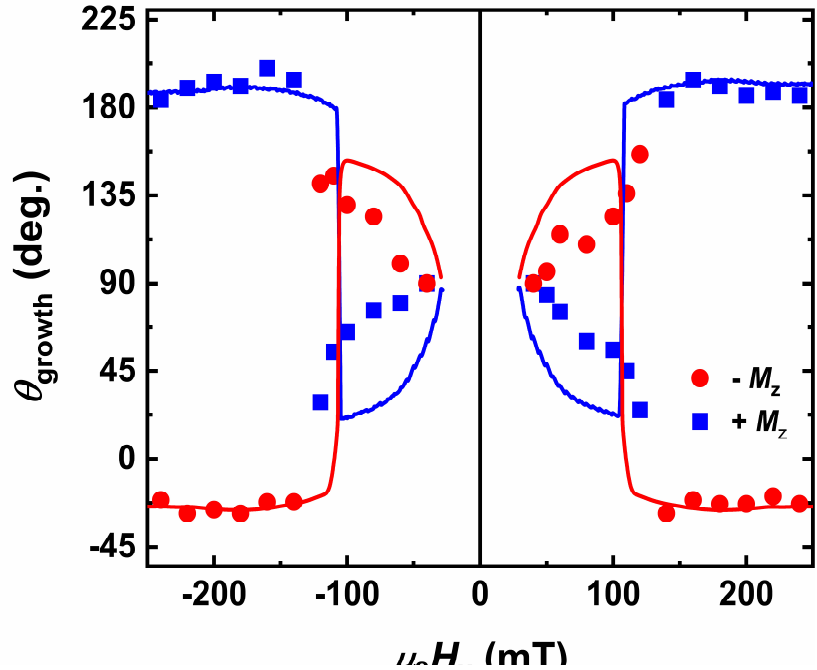

**Figure S4:** The growth direction $\theta_{growth}$ of both "up" ($+M_z$) and "down" ($-M_z$) dendritic domains as a function of $\mu_0 H_x$, determined both experimentally (symbols) and analytically with steady-state dynamics and dispersive stiffness (lines) for the [Co(0.7 nm)/Ni(0.5 nm)/Pt(0.7 nm)]$_3$ sample. When plotting the analytically predicted $\theta_{growth}$ profiles, we only display data for in-plane fields where a > 60 % growth asymmetry between the fastest and slowest moving azimuthal positions is predicted (*i.e.*, $|\mu_0 H_x| \geq 40$ mT). The angular convention used to define the domain growth direction $\theta_{growth}$ relative to the in-plane field $\mu_0 H_x$ direction is defined in Figure 2b of the main text.

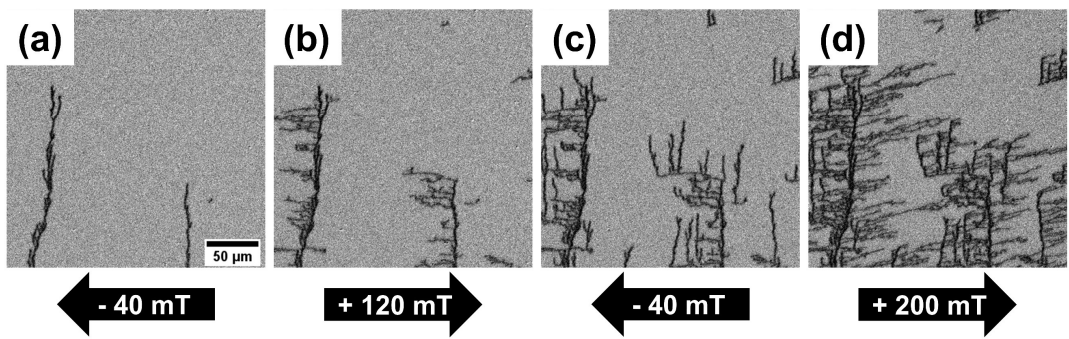

**Supplemental Figure S5:** Polar MOKE images of the [Co(0.7 nm)/Ni(0.5 nm)/Pt(0.7 nm)]$_3$ sample as 5 ms-long, +15 mT perpendicular magnetic field pulses were applied in static in-plane fields of (a) - 40 mT, (b) + 120 mT, (c) - 40 mT, and (d) + 200 mT. The initial dendrite state was obtained by applying several 5 ms-long, +15 mT perpendicular field pulses in a static in-plane field of - 40 mT.

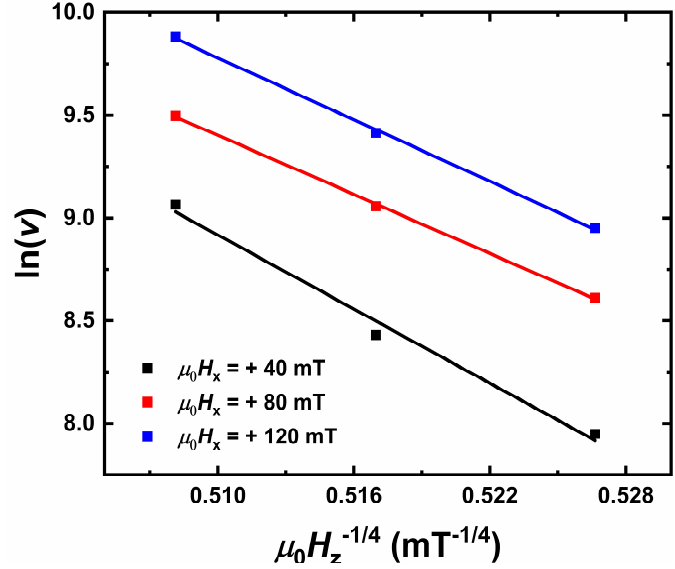

**Figure S6:** Verification of the creep scaling law in the [Co(0.7 nm)/Ni(0.5 nm)/Pt(0.7 nm)]$_3$ sample, obtained from measurements of the expansion velocity $v$ (in µm/s) for different perpendicular fields $\mu_0 H_z$ and in-plane fields $\mu_0 H_x$. The data was collected using a $+M_z$ domain, subjected to 5 ms-long $\mu_0 H_z$ pulses. Given the strong growth asymmetry observed in the sample, the velocity was determined at the azimuthal position that exhibits the fastest expansion for each $\mu_0 H_x$.

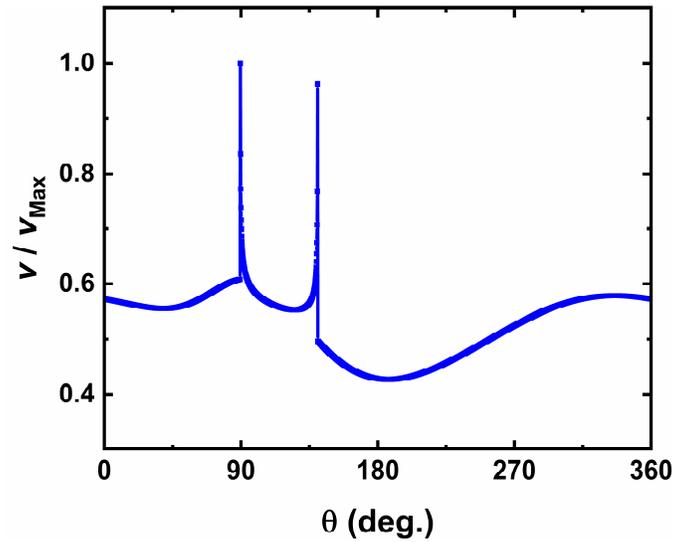

**Figure S7:** Relative velocity $v/v_{\text{Max}}$ as a function of azimuthal position $\theta$ for $+M_z$ dendritic stripe domains in the [Co(0.7 nm)/Ni(0.5 nm)/Pt(0.7 nm)]$_3$ sample in an applied in-plane field $\mu_0 H_x = +$ 37 mT, calculated using the long-wavelength ($L \to \infty$) dispersive stiffness treatment. $\theta$ is defined using the angular convention shown in Figure 2b of the main text.

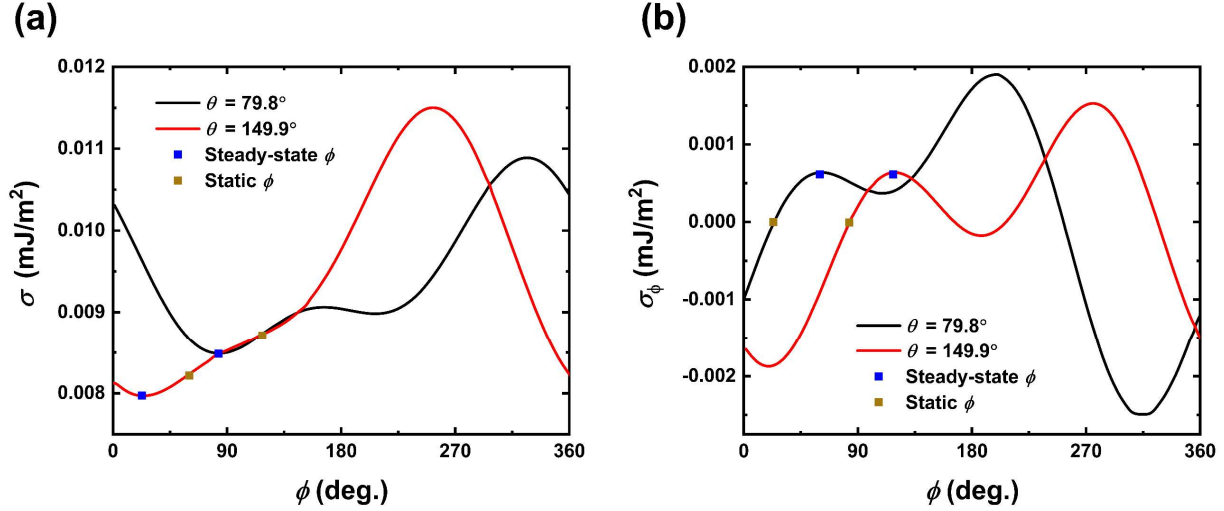

**Figure S8:** Plots showing the $\varphi$-dependence of the **(a)** calculated domain wall energy $\sigma$ and **(b)** normalized restoring torque $\sigma_\varphi$ at azimuthal positions of $\theta = 80°$ (black) and $\theta = 150°$ (red) for the [Co(0.7 nm)/Ni(0.5 nm)/Pt(0.7 nm)]$_3$ sample. These wall positions correspond to the discontinuities present in the $\varphi(\theta)$ profiles for $\mu_0 H_x = +40$ mT. The $\varphi$ solutions for the static (gold squares) and steady-state (blue squares) treatments are labeled for each $\theta$, using the angular definition provided in Figure 2b of the main text. The full $\sigma(\varphi)$ profiles for the static and steady-state cases for this $\mu_0 H_x$ are shown in Figure 3b of the main text.

### Section S2. Velocity asymmetry and calculation of iDMI energy density

The velocity $v$ of "↑ to ↓" and "↓ to ↑" DWs as a function of in-plane applied field $\mu_0 H_x$ was determined for the [Co(0.7 nm)/Ni(0.5 nm)/Pt(0.7 nm)]$_2$ sample using the process employed by Je et al.[2], and later used by many others. These velocity curves are shown in Figure S9. In Figure S9, each type of DW exhibits a respective minimum in velocity depending on the direction in which $\mu_0 H_x$ is applied. By calculating the field at which this minimum in velocity occurs ($\mu_0 H_{DMI}$), the iDMI energy density ($D_{DMI}$) can be determined using the expression:

$$D_{DMI} = \mu_0 H_{DMI} M_S \sqrt{\frac{A}{K_{eff}}} \quad . \tag{S1}$$

All parameters in Equation S1 have previously been defined in the main text. From the data in Figure S9, we have extracted a $H_{DMI}$ of 120 mT, which for the material parameters of the [Co(0.7 nm)/Ni(0.5 nm)/Pt(0.7 nm)]$_2$ sample corresponds to $D_{DMI} = -0.63$ mJ/m$^2$. The left-handed $D_{DMI}$ is inferred from the configuration (i.e., ↑ to ↓ versus ↓ to ↑) of the DWs and the $\mu_0 H_x$ direction.

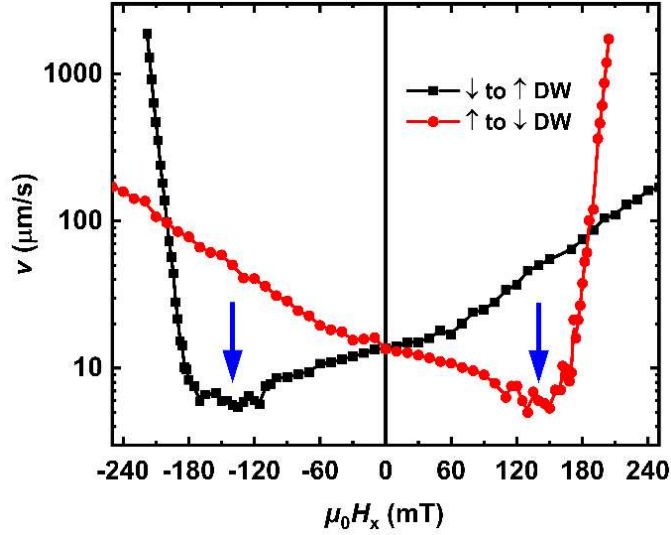

**Figure S9.** Expansion velocity as a function of in-plane magnetic field $\mu_0H_x$ of ↑ to ↓ and ↓ to ↑ domain walls in the [Co(0.7 nm)/Ni(0.5 nm)/Pt(0.7 nm)]$_2$ sample. Data was collected using 5 ms-long perpendicular field pulses of $\mu_0H_z = +15$ mT. Blue arrows indicate the commonly-attributed measure of the effective magnetic field generated by the iDMI ($\mu_0H_{DMI}$).

**Section S3. Long-wavelength dispersive stiffness in the [Co/Ni/Pt]$_2$ sample**

In Figure S10, we plot the 1D elastic energy $\sigma$ and the long-wavelength ($L \to \infty$) stiffness $\tilde{\sigma}$ of a $+M_z$ domain in the [Co(0.7 nm)/Ni(0.5 nm)/Pt(0.7 nm)]$_2$ sample as a function of in-plane applied magnetic field strength, calculated using the expression[1]:

$$\tilde{\sigma}(\theta) = \sigma_{eq}(\theta) + \frac{\partial^2 \sigma_{eq}}{\partial \theta^2}(\theta) \tag{S2}$$

where $\sigma_{eq}$ refers to the energy profiles calculated using Equation 1 in the main text. In Figure S10, we consider the situations in which the domain wall normal is parallel ($\theta = 0°$) and opposite ($\theta = 180°$) to the applied in-plane field axis. Under the treatment of elastic stiffness in the long-wavelength regime, the onset of "unphysical" stiffness values roughly corresponds to the $\mu_0H_x$ values at which a reversal in circular domain growth direction was experimentally found to occur in the [Co(0.7 nm)/Ni(0.5 nm)/Pt(0.7 nm)]$_2$ sample (main text Figure 1a-d).

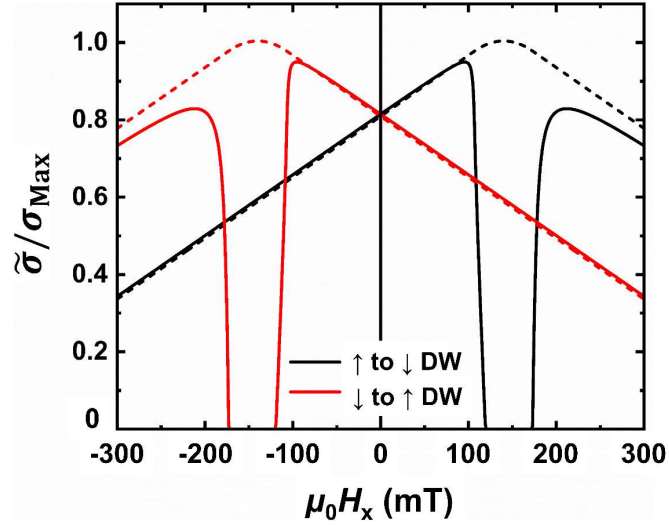

**Figure S10.** The long-wavelength stiffness $\tilde{\sigma}$ (solid lines) and 1D linear DW energy density $\sigma$ (dashed lines) of a domain wall as a function of in-plane applied magnetic field $\mu_0 H_x$ for ↑ to ↓ and ↓ to ↑ domain walls [Co/ Ni/ Pt]$_2$ sample. The $\tilde{\sigma}$ values have been normalized to the resting energy of a Bloch domain wall $\sigma_0$.

## Supporting References